\documentclass[preprint2]{aastex}

\usepackage{graphicx}
\usepackage{epstopdf}
\usepackage{epsfig}

\shorttitle{DASCH Light Curves of Kepler Planet Host Stars}
\shortauthors{Tang et al.}

\begin{document}


\title{100-year DASCH Light Curves of Kepler Planet-Candidate Host Stars}


\author{Sumin Tang\altaffilmark{1, 2, 3}, Dimitar Sasselov\altaffilmark{1}, Jonathan Grindlay\altaffilmark{1},
Edward Los\altaffilmark{1}, Mathieu Servillat\altaffilmark{1,4}}
\altaffiltext{1}{Harvard-Smithsonian Center for Astrophysics, 60
Garden St, Cambridge, MA 02138}
\altaffiltext{2}{Kavli Institute for Theoretical Physics, University of California, Santa Barbara, CA 93106}
\altaffiltext{3}{Division of Physics, Mathematics, \& Astronomy, California Institute of Technology, Pasadena, CA 91125, USA}
\altaffiltext{4}{Laboratoire AIM (CEA/DSM/IRFU/SAp, CNRS, Universite Paris Diderot),
 CEA Saclay, Bat. 709, 91191 Gif-sur-Yvette, France}


\begin{abstract}
We present 100 year light curves of Kepler planet-candidate host stars
from the Digital Access to a Sky Century at Harvard (DASCH) project.
261 out of 997 host stars
have at least 10 good measurements on DASCH scans of the Harvard plates.
109 of them have at least 100 good measurements, including 70\%  (73 out of 104) of all host stars with $g\leq13$ mag,
and 44\% (100 out of 228) of all host stars with $g\leq14$ mag.
Our typical photometric uncertainty is $\sim0.1-0.15$ mag.
No variation is found at $3\sigma$ level for these host stars,
including 21 confirmed or candidate hot Jupiter systems which might be expected
to show enhanced flares from magnetic interactions between dwarf
primaries and their close and relatively massive planet companions.
\end{abstract}


\keywords{Planetary Systems - Stars: general}

\section{Introduction}

In 2011 the Kepler Mission announced 997 stars with a total of 1235
planetary candidates that show transit-like signatures (Borucki et al. 2011).
It is estimated that more than 95\% of these candidates are planets (Morton \& Johnson 2011),
although a much higher false positive rate of 35\% for Kepler close-in giant candidates is reported by Santerne et al. (2012).
Kepler has completed 3 years of observations and with a 4-year extension will accumulate 7.5 years of data
for these host stars. Searching the host stars for significant very long-term photometric variations on timescales
of decades and for rare flare events is important for understanding fully the environments of their
planets and for constraining the habitability of exoplanets in general.
For example, a recent study by Lecavelier des Etangs et al. (2012) found increased evaporating atmosphere of hot Jupiter HD 189733b during a transit,
which is probably related to a X-ray flare 8h before the transit.
Here we present the 100-year light curves of Kepler planet-candidate host stars
from the Digital Access to a Sky Century at Harvard (DASCH) project (Grindlay et al. 2009, 2012).

The photometric variability of about 150,000 Kepler target stars, including the sample with planetary
candidates, shows low-level variability due to modulation by spots and other
manifestations of magnetic activity, e.g., white light flares (Basri et al. 2011).
None of these variabilities have amplitudes that would be detectable ($>0.3$ mag if only a single observation) in the DASCH data.
Here we aim to complement
the short-timescale and high-precision photometry by Kepler with the much longer timescales available from DASCH.
The DASCH project can reveal stars with rare flare events or extremely slow changes in luminosity.
It is of interest to search for both kinds of variability with DASCH. Extremely large ($>$1mag) flares
could drive significant mass loss from the atmospheres of hot Jupiters,
and long term changes in host star luminosity could drive changes in planetary winds and spectral composition.
This paper provides DASCH light curves
for the initial sample of Kepler planets presented by Borucki et al. (2011).

\section{DASCH light curves}
In order to take advantage of the unprecedented Kepler data on short timescales (Borucki et al. 2010),
which complements DASCH data on long timescales,
we have scanned and processed $3735$ plates taken from the 1880s to 1990 in or covering part of the Kepler field.
Each plate covers 5$-$40 degrees on a side, and most of them are blue-sensitive (close to Johnson B).
Most plates have limiting magnitudes $12-14$ mag, while $\sim9\%$ of them (mc series) are down to $\sim17$ mag.
More details on the coverage and limiting magnitudes of the plates in the Kepler field are described in Tang et al. (2013).

We used the Kepler Input Catalog (KIC; Brown et al. 2011) for photometric calibration.
The typical relative photometric uncertainty is $\sim0.1-0.13$ mag,
as measured by the median light curve RMS of stars (Laycock et al. 2010; Tang et al. 2013; see also Table 1 and Figure 3 in this paper).
Given that most stars are constant at the $0.1$ mag level,
their light curve RMS values are dominated by photometric uncertainty,
and thus the median light curve RMS represents the relative photometric uncertainty.
The typical absolute photometric uncertainty, as measured by
the difference between our measurements and KIC g band magnitudes,
is $\sim0.2$ mag, and up to $\sim0.35$ mag for bright stars ($g<11$ mag).
The larger absolute photometric uncertainty is caused by additional contributions from the
uncertainties in KIC g mag (especially at the bright end),
and the difference between the effective color of the plates (close to Johnson B) and g band.
The magnitude uncertainties we provide in DASCH light curves,
are defined as the scattering of ($m_{DASCH}-g_{KIC}$) for stars in local spatial bins with similar magnitudes,
and thus represent the absolute photometric uncertainty (Tang et al. 2013).
The typical absolute astrometric uncertainty (per data point) is $\sim0.8-5''$,
depending on plate scale (Laycock et al. 2010; Los et al. 2011; Servillat et al. 2011).

Example DASCH light curves of 3 Kepler hot Jupiter host stars, as well as a binary are shown in Figure 1.
Fainter stars have fewer DASCH measurements, due to smaller number of deeper plates available.
For a given star, every measurement is typically separated by days to months, and could be up to years.
Therefore, one stellar flare, if happened during the plate observation, is expected to appear on a single plate only.
Some fields do have multiple exposures which our pipeline is able to process individually (see Los et al 2011),
and would allow stellar flares to be confirmed by successive measurements; no such flares were seen.
Among the 4 stars plotted in Fig. 1, K10666592 (HAT-P-7; P\'{a}l et al. 2008), K10264660 (Kepler-14; Buchhave et al. 2011)
and K8191672 (Kepler-5; Koch et al. 2010) are confirmed hot Jupiter host stars. 
K5122112 (KOI 552; Borucki et al. 2011) was listed as a hot Jupiter candidate, but later has been unveiled as a binary (Bouchy et al. 2011).
All three planets have $R\geq R_{Jupiter}$, $a<0.1$ AU, and equilibrium temperature $>1300$ K.
No variation is detected.

\begin{figure} [tb]
\epsfig{file=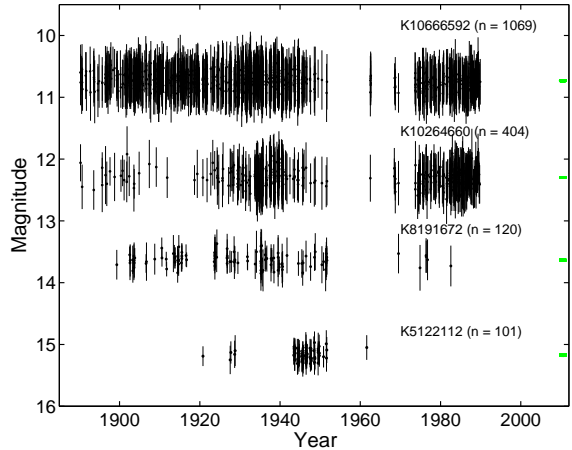, angle=0, width=\linewidth}
\caption{Example DASCH light curves of 3 Kepler hot Jupiter host stars, and a binary.
From top to bottom: KIC 10666592 (HAT-P-7; P\'{a}l et al. 2008), 10264660 (Kepler-14; Buchhave et al. 2011), 8191672 (Kepler-5; Koch et al. 2010), and 5122112 (KOI 552, binary; Bouchy et al. 2011).
Number of measurements ($n$) in each light curve are also shown, which drops for fainter stars due to smaller number of deeper plates available.
Kepler PDC corrected light curves are also shown in green dots for comparison;
note that the variation in the Kepler light curves for these stars are at $10^{-4}$ to $10^{-3}$ level, much smaller than the plotting symbols.
The DASCH results constrain any rare but bright ($>0.4$ mag) flares from Kepler planet host stars, which might be most likely from hot Jupiter systems.
}
\end{figure}

All the light curve data and plots are available at the DASCH website\footnote{http://hea-www.harvard.edu/DASCH/papers/Kepler-planet-host-star-lightcurves}.
Only good measurements are included.
We excluded blended images,
measurements within $0.75$ mag of the limiting magnitude which are more likely to be contaminated by noise,
images within the outer border of the plates whose width is 10\% of the plate's minor-axis length (annular bin 9; see Laycock et al 2010),
and dubious points with image profiles different from neighbor stars and thus are suspected to be emulsion defects or dust.
Stars with strong correlation between magnitude measurements and plate limiting magnitudes,
or between magnitude measurements and plate astrometry uncertainties,
are also excluded, which are very likely to be polluted by noise or blends.
More detailed descriptions can be found at Tang et al. (2013).

Note that these Kepler planet host stars have variations at the $0.01\%-0.1\%$ level in the Kepler light curves,
which are much smaller than the plotting symbols in Figure 1 and our photometric accuracy.
For comparison, the Kepler light curves of the 4 example Kepler planet-candidate host stars (from Q0 or Q1 to Q6)
are also shown in Figure 1 as green dots.
We used the PDC corrected flux, which are converted to magnitudes and shifted to the mean magnitudes of
DASCH light curves (See http://keplergo.arc.nasa.gov/CalibrationSN.shtml).

Among 997 host stars, 261 stars have at least 10 good measurements on DASCH plates,
and 109 stars have at least 100 good measurements.
Distributions of g band magnitudes for all the host stars and host stars with at least 10 or 100 DASCH measurements,
are shown in Figure 2.
We have at least 100 measurements for 70\%  (73 out of 104) of all host stars with $g\leq13$ mag,
and 44\% (100 out of 228) of all host stars with $g\leq14$ mag.
Most stars brighter than $g=13$ mag we lost (i.e. with less than 100 good measurements),
were due to blending with neighbor stars, as expected in such a crowded field,
with 74\% of them (23 out of 31) having bright neighbor stars with $g<14$ mag within 1 arcmin.
For comparison, for $g\leq13$ stars with at least 100 good measurements, only 14\% (10 out of 73) of them have $g<14$ mag neighbor stars within 1 arcmin.

\begin{figure} [tb]
\epsfig{file=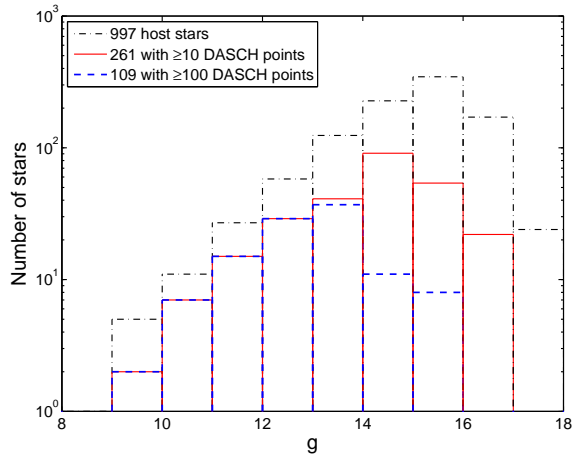, angle=0, width=\linewidth}
\caption{Distribution of g band magnitudes for 997 Kepler planet-candidate host stars from Borucki et al. 2011 (black dash-dotted line),
261 host stars with at least 10 DASCH measurements (red solid line),
and 109 host stars with at least 100 DASCH measurements (blue dashed line).}
\end{figure}

We have carefully examined light curves of the 261 planet candidate host stars,
and none of them showed variations (flares or dips) at the $>3\sigma$ level,
where $\sigma$ here is our absolute photometric uncertainty with typical value of $\sim0.2$ mag for $g>13$ objects,
and up to $0.35$ mag for $g<12$ objects.
Nor did we see any long-term photometric trends.
We also compare the light curve RMS of these host stars vs. other stars with similar magnitudes in the Kepler field,
as shown in Figure 3.
The raw light curve RMS vs. g band magnitude for the 261 host stars with at least 10 DASCH measurements are shown as blue dots.
The black solid line shows the median light curve RMS of all the stars with at least 10 DASCH measurements in the Kepler field,
and the red dashed lines and green dash-dotted lines show the $1\sigma$ and $2\sigma$ distributions, respectively.
The median, $1\sigma$, and $2\sigma$ distributions of light curve RMS are calculated in 0.5 magnitude bins,
after three iterations of $4\sigma$-clipping, to exclude variable stars and dubious light curves contaminated by blending or plate defects.
Stars with bright neighbors ($\geq30\%$ the flux of the star in KIC g band) within 30'' are also excluded in the calculation of median, $1\sigma$, and $2\sigma$ distributions,
to avoid contamination from blending.
Note the drop of RMS for stars fainter than 14 mag is due to the fact that in general deeper plates are of better quality.
Our plates are not a homogeneous sample. Stars with different magnitudes, even in the same region of the sky, are actually covered by different plates.
For example, a 12th mag star is detected on all the plates deeper than 12th mag, while a 15th mag star is only detected on plates deeper than 15th mag.
Most of the deeper plates, such as mc series, are of much better quality compared with the shallow plates,
which leads to the decrease of typical (median, etc.) light curve rms of fainter stars (g$>$14 mag).

\begin{figure} [tb]
\epsfig{file=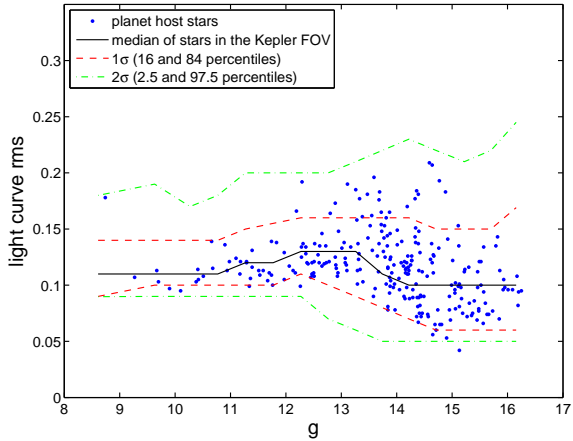, angle=0, width=\linewidth}
\caption{DASCH light curve RMS vs. g band magnitude.
The 240 Kepler planet-candidate host stars with at least 10 DASCH measurements are plotted as blue dots.
Light curve RMS percentiles of 59,453 stars with at least 10 DASCH measurements and no bright neighbors in the Kepler field are shown for comparison;
stars with bright neighbors ($\geq30\%$ the flux of the star in KIC g band) within 30'' are excluded to avoid contamination from blending.
The 50th percentile (median) is shown in black solid line,
the 16th and 84th percentiles ($1\sigma$) are shown in red dashed line,
and the 2.5th and 97.5th percentiles are shown in green dash-dotted lines.
The median, $1\sigma$, and $2\sigma$ distributions of light curve RMS are calculated in 0.5 magnitude bins,
after three iterations of $4\sigma$-clipping, to exclude variable stars and dubious light curves contaminated by blending or plate defects.
}
\end{figure}

None of the planet candidate host stars has a light curve RMS more than $2\sigma$ greater than the median RMS of stars of similar magnitudes.
There is one planet host star, i.e. Kepler-21 (KIC 3632418, $g=8.74$; Howell et al. 2012), with light curve RMS close to the $2\sigma$ distribution.
Further examination shows that some images are marginally contaminated by a $g=11.43$ mag neighbor star located at 51'' away from the star,
and thus its RMS excess is dubious.

It has been suggested that the magnetic interaction between hot Jupiters and their host stars should enhance stellar activity,
and may lead to phase shifts of hot spots on the stellar chromosphere
(Cuntz \& Shkolnik 2002; Shkolnik et al. 2003, 2005; Kopp et al. 2011; Poppenhaeger \& Schmitt 2011).
Given the extremely long timescale covered by DASCH, it is interesting to examine the light curves of stars hosting hot Jupiters.
Among the 261 planet candidate host stars with at least 10 DASCH measurements,
21 of them host hot Jupiter planet candidates ($R\geq R_{Jupiter}$ and $a<0.4$ AU),
and 9 of them have at least 100 DASCH measurements (4 are shown in Figure 1 as examples).
DASCH light curve properties of these Jupiter planet candidate host stars,
including number of measurement, median magnitude and RMS in the light curves,
are listed in Table 1.
None of them showed variations at the $>3\sigma$ level.

\section{Conclusion}
The Kepler mission has now discovered more than 2000 planetary candidates (only the first 1235 considered here) and provided unparalleled precision
light curves for their host stars. Here we complement that database with much longer timescale 100-year
light curves from the DASCH project. Despite their inferior photometric accuracy, the DASCH light curves
sample such an extended period of time (e.g., tens of solar-like cycles of activity),
that rare or very slow phenomena can be studied.
From the statistical sample of the Kepler planet host stars, limits on their long-term variations
and rare flare events, such as the X-ray/EUV flare from HD189733 (Lecavelier des Etangs et al. 2012),
help us understand the planetary environments around main sequence stars
and the habitability of exoplanets in general.
We note that the HD189733 system is not within the Kepler field and
so has not yet been scanned by DASCH.
However, given the luminosity of the flare ($7\times10^{28}$ erg s$^{-1}$; Lecavelier des Etangs et al. 2012),
and the relatively luminous (K1.5V) host star, such a flare is beyond the detection limit of DASCH.
Relatively larger optical flares
would be expected for cooler stars with hot Jupiter companions.
Because Kepler is mostly targeting on Sun like stars,
none of M dwarf and hot Jupiter system is in the sample of the 261 host stars we studied.

We have scanned and processed $3735$ plates taken from the 1880s to 1990 in or covering part of the Kepler field,
and studied the light curves of 261 planet host stars that have at least 10 good measurements on DASCH plates.
We find no photometric variations at the $3\sigma$ level. 
All the light curve data and plots of planet hosts in this study are available at
the DASCH website\footnote{http://hea-www.harvard.edu/DASCH/papers/Kepler-planet-host-star-lightcurves}.
Besides, we have released all the DASCH data in the Kepler field to the public in DASCH Data Release 1\footnote{http://dasch.rc.fas.harvard.edu}.
DASCH light curves over $\sim$100 year timescales will continue to provide
unique constraints for planet host stars, 
as well as any other interesting objects in the Kepler field.

\acknowledgments
We thank the anonymous referee for suggestions that have helped improve this paper.
We thank Alison Doane, Jaime Pepper, David Sliski and Robert J. Simcoe at CfA for their work on DASCH,
and many volunteers who have helped digitize logbooks, clean
and scan plates (http://hea-www.harvard.edu/DASCH/team.php).
This work was supported in part by NSF grants AST0407380 and AST0909073
and now also the \emph{Cornel and Cynthia K. Sarosdy Fund for DASCH}. \\


\clearpage

\begin{table*}[tb]
\caption{DASCH light curve properties of host stars of confirmed and candidate hot Jupiter planets in the Kepler field.} \centering
\begin{minipage}{\textwidth}
\tabcolsep 7pt
\begin{tabular}{llrrrrc}
\tableline
KOI & KIC	&	KIC g &	$N_{lc}$\tablenotemark{1}	&	lc median\tablenotemark{2}	&	lc RMS\tablenotemark{3}	 & Notes \tablenotemark{4} \\
	&       & (mag)	&	 &	(mag)	&	(mag)	&  \\
\tableline
1	&	11446443	&	11.74	&	570	&	11.72	&	0.109	&	TrES-2; O'Donovan et al. (2006)	\\
2	&	10666592	&	10.94	&	1069	&	10.73	&	0.113	&	HAT-P-7; P\'{a}l et al. (2008)	\\
13	&	9941662	&	9.7	&	1348	&	9.76	&	0.103	&	Kepler-13; Shporer et al. (2011)	\\
18	&	8191672	&	13.88	&	120	&	13.63	&	0.099	&	Kepler-5; Koch et al. (2010)	\\
20	&	11804465	&	13.78	&	62	&	13.68	&	0.153	&	Kepler-12; Fortney et al. (2011)	\\
97	&	5780885	&	13.26	&	170	&	13.13	&	0.185	&	Kepler-7; Latham et al. (2010)	\\
98	&	10264660	&	12.36	&	404	&	12.3	&	0.128	&	Kepler-14; Buchhave et al. (2011)	\\
100	&	4055765	&	12.91	&	243	&	12.85	&	0.115	&		\\
128	&	11359879	&	14.27	&	32	&	14.185	&	0.122	&	Kepler-15; Endl et al. (2011)	\\
135	&	9818381	&	14.35	&	38	&	14.29	&	0.101	&	Kepler-43; Bonomo et al. (2012)	\\
138	&	8506766	&	14.22	&	49	&	14.14	&	0.113	&		\\
191	&	5972334	&	15.56	&	54	&	15.635	&	0.066	&		\\
203	&	10619192	&	14.64	&	15	&	14.52	&	0.207	&	Kepler-17; D\'{e}sert et al. (2011)	\\
846	&	6061119	&	16.01	&	45	&	16.02	&	0.096	&		\\
897	&	7849854	&	15.82	&	13	&	15.86	&	0.143	&		\\
976	&	3441784	&	9.9	&	1407	&	9.73	&	0.097	&		\\
1020	&	2309719	&	13.34	&	84	&	13.14	&	0.114	&		\\
1452	&	7449844	&	13.84	&	79	&	13.81	&	0.116	&		\\
1474	&	12365184	&	13.28	&	242	&	13.255	&	0.158	&		\\
1540	&	5649956	&	16.17	&	45	&	16.36	&	0.108	&		\\
1549	&	8053552	&	15.73	&	10	&	15.725	&	0.077	&		\\
\tableline
\end{tabular}
\tablenotetext{1}{Number of measurements in DASCH light curve.}
\tablenotetext{2}{Median magnitude in DASCH light curve.}
\tablenotetext{3}{DASCH light curve RMS.}
\tablenotetext{4}{Names of the confirmed hot Jupiter systems and references.}
\end{minipage}
\end{table*}

\end{document}